# Iron and Nickel spectral opacity calculations in conditions relevant for pulsating stellar envelopes and experiments

D. Gilles [1,a], S. Turck-Chièze [1], M. Busquet [2], F. Thais [3], G. Loisel [4], L. Piau [1], J.E. Ducret [1], T. Blenski [3], C. Blancard [5], P. Cossé [5], G. Faussurier [5], F. Gilleron [5], J.C. Pain [5], Q. Porcherot [5], J.A. Guzik [6], D.P. Kilcrease [6], N.H. Magee [6], J. Harris [7], S. Bastiani-Ceccotti [8], F. Delahaye [9], C.J Zeippen [9]

[1]CEA/IRFU/Sap, F-91191 Gif-sur-Yvette, Cedex, France
[2]ARTEP Ellicott City, MD 21042, USA
[3]CEA/IRAMIS/SPAM, F-91191 Gif-sur-Yvette, France
[4]Sandia National Laboratories, Albuquerque, NM 87185-1196, USA
[5]CEA/DIF, F-91297 Arpajon, France
[6]Theoretical Division, LANL, Los Alamos NM 87545, USA
[7]AWE Readings Berkshire, RG7 4PR, UK
[8]LULI, Ecole Polytechnique, 91128 Palaiseau, France
[9]LERMA Observatoire de Paris-Meudon, France

**Abstract.** Seismology of stars is strongly developing. To address this question we have formed an international collaboration OPAC to perform specific experimental measurements, compare opacity calculations and improve the opacity calculations in the stellar codes [1]. We consider the following opacity codes: SCO, CASSANDRA, STA, OPAS, LEDCOP, OP, SCO-RCG. Their comparison has shown large differences for Fe and Ni in equivalent conditions of envelopes of type II supernova precursors, temperatures between 15 and 40 eV and densities of a few mg/cm$^3$ [2, 3, 4]. LEDCOP, OPAS, SCO-RCG structure codes and STA give similar results and differ from OP ones for the lower temperatures and for spectral interval values [3]. In this work we discuss the role of Configuration Interaction (CI) and the influence of the number of used configurations. We present and include in the opacity code comparisons new HULLAC-v9 calculations [5, 6] that include full CI. To illustrate the importance of this effect we compare different CI approximations (modes) available in HULLAC-v9 [7]. These results are compared to previous predictions and to experimental data. Differences with OP results are discussed.

## 1 Theoretical opacity spectra for pulsating stellar envelope conditions

In recent papers [2, 3] we have compared a large number of opacity calculations for experimental conditions relevant for stellar envelope conditions of intermediate mass stars. The temperatures are between 15 and 40 eV and densities of a few mg/cm$^3$. The codes are based on different approaches:

[a] email : dominique.gilles@cea.fr

statistical (SCO, STA), detailed (OPAS, OP, LEDCOP) or mixed (SCO-RCG) [2, 3, 4]. These comparisons show some differences in the frequency-dependent spectra among all codes but at low temperature and in the spectral range of our comparisons indicate a clearly distinct behavior of the OP spectra with respect to all other codes. This is illustrated by Fig. 1a (Iron, T= 27.3eV and ρ= 3.4mg/cm3) and Fig. 1b (Nickel, T= 15.3eV and ρ= 5.5mg/cm3).

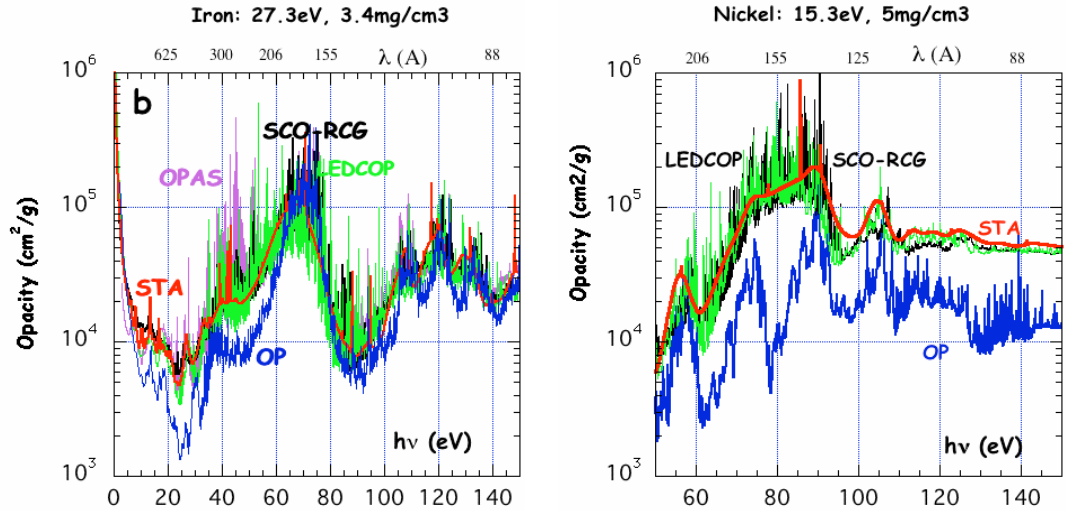

**Fig. 1a (Fe), 1b (Ni).** Opacity spectra from LEDCOP, OPAS, STA & SCO-RCG in the spectral range important for the calculation of the mean Rosseland values generally used in stellar physics.

As the OPAS, SCO-RCG and STA results remain comparable, two questions appear. The first one is the role of the general configuration interaction (CI) and the second one is the set of configurations included in the OP calculations used in stellar evolution codes. This second question is related to the difficulty of taking into account numerous excited levels in the extended database of astrophysics.

In this study, the role of CI and the influence of the number of excited levels is discussed, thanks to the new version of HULLAC-v9 code with full CI calculations. In a companion paper, definitions and illustrations of these calculations are given [7]. In this paper we compare also HULLAC-v9 predictions to two experimental conditions 1/ Da Silva et al (iron) [8, 9] and 2/ LULI 2010 experiment (nickel) first analysis [2].

## 2 Configuration Interaction (CI) treatments in theoretical codes [4]

In HULLAC-v9 and OP calculations full Configuration Interaction (CI) can be included [5, 6, 10]. In HULLAC-v9, the computation of fine structure levels with CI (mode L for Levels) is chosen by the user for defined groups of levels (GRL), see [7]. Selecting the "CIinNRC" mode, HULLAC-v9 computes the usual Relativistic CI (RCI) between all levels of one Non Relativistic Configuration (NRC), see [5, 6] for more details or illustrations. The diagonalization of eigenvectors for all levels of same J is performed in a GRL. It is also possible to turn off the CI calculation using sub mode "NoCI". Other GRL can be selected, like Layzer complexes (configurations with identical n-shell occupancies). In OP full CI is included but the difficulty is to take into account numerous excited levels in the R-Matrix formalism. STA, OPAS and SCO-RCG codes account for RCI. The statistical code STA computes Unresolved Transition Arrays (UTA) or Super Transition Arrays (STA), but introduces corrections for positions and intensities of the UTA for the RCI effects. Results appear as envelopes of the resolved lines or spectrum. SCO-RCG is a "hybrid" opacity code that combines the statistical Super-Transition-Array (STA) approach and fine-structure calculations for relevant STA's.

It relies on some criteria to decide whether a detailed treatment of the lines is necessary or if statistical approach can be used. It then uses either RCG routines from the Cowan's code or the UTA/STA formalism from the SCO code.

## 3 Iron: comparison with transmission spectrum at 25eV and 8 mg/cm$^3$

The transmitted measurement of Da Silva et al. [8, 9] covers the spectral range from ≈ 50 to 120 eV where the opacity is dominated by contributions from $\Delta n= 0$, n= 3 iron transitions below 80 eV and n= 3 -> n= 4 transitions ("3L-4L") above 80 eV. This sub-keV spectral region is known to be dominant for the Rosseland mean opacity. We have compared full CI HULLAC-v9 calculations, including the 3L-4L transitions, to ab initio OP and OPAL predictions at 25eV shown in [9], STA results and Da Silva experimental results (Fig. 2a). A quite good agreement is found at these conditions between STA, HULLAC-v9 and OP results. OPAL results are significantly shifted. The influence of the number of excited levels (3L-4L transitions of ions Fe VIII to Fe X, for example) and of CI effect (mode Level "L") is illustrated in Fig. 2b where 3 HULLAC-v9 transmission spectra are also compared to Da Silva experimental results. Comparing Fig. 2a and Fig. 2b we can say that OP and HULLAC-v9 results are comparable even if not totally the same at these conditions.

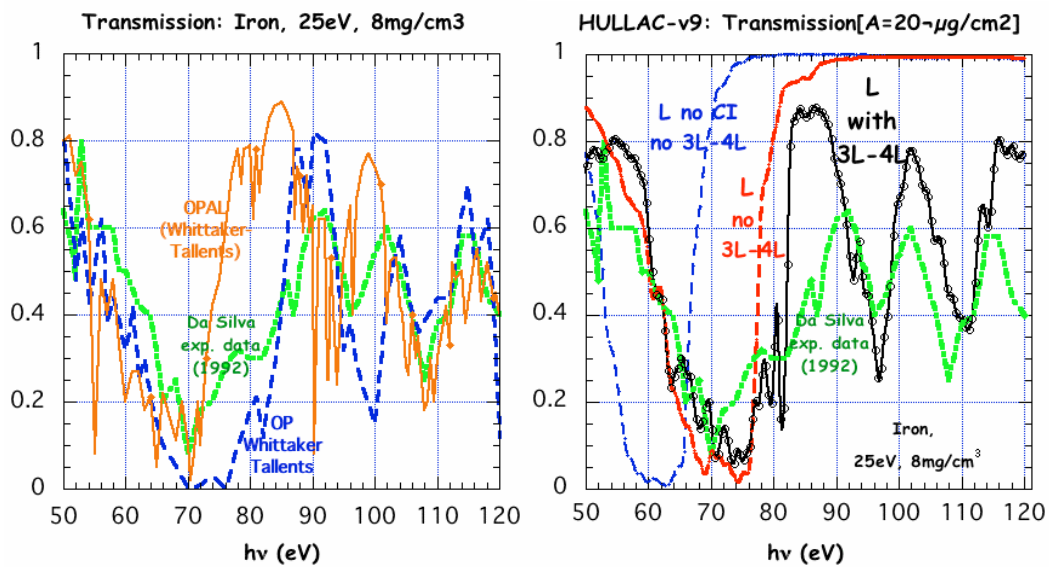

**Fig. 2a, b. a/** Da Silva transmission results compared to OPAL and OP transmissions (A= 20 μ g/cm2). This figure is derived from [9]. **b/** HULLAC-v9 transmissions (A= 20 μg/cm2) are calculated with or without CI and 3L-4L transitions (red dashed line: full CI mode "L" without 3L-4L transitions, dotted black line: full CI mode "L" with 3L-4L transition, solid blue line mode L without CI "LnoCI" and without 3L-4L transitions. Inclusion of both CI and 3L-4L transitions is crucial in the calculation (Transmission= exp (-opacity(cm2/g)*A(g/cm2))).

## 4 Nickel: Conditions of the LULI 2010 experiment

For nickel at T=15.3 eV the number of lines and of CI coefficients yields to dramatic increase of the computation time. Complete Nickel HULLAC-v9 full CI calculations including the 3L-4L transitions are still in progress. So only major contributions coming from $\Delta n=0$, n=3 transitions have been reported on Figure 3. Nevertheless we confirm with these HULLAC-v9 results, that large differences exist between atomic codes (represented here by STA for clarity, red line) and OP monochromatic nickel opacity results (blue line) for energies between 70 and 95 eV. The Nickel OP opacities presented here are extracted from the OPCD database [4]. HULLAC-v9 opacities (dashed line) are calculated in full CI mode for the transition (Δn=0, n=3) of importance in the spectral range of the measured data. The dip in the OP opacity around 80 eV can be due to the difficulty of taking into account numerous excited levels at these conditions. Nevertheless the agreement is good for lower energies (left part of Fig. 3). The asymptotic behavior of the OP opacity values for high energies are not in agreement with those of all other codes (right part of Fig. 3), though nickel opacities play little role in solar mixtures because this element is not abundant. In the present case partial CI are found to give a good description of the spectrum (as in STA calculation), but accounting for a large number of configurations appears to be crucial.

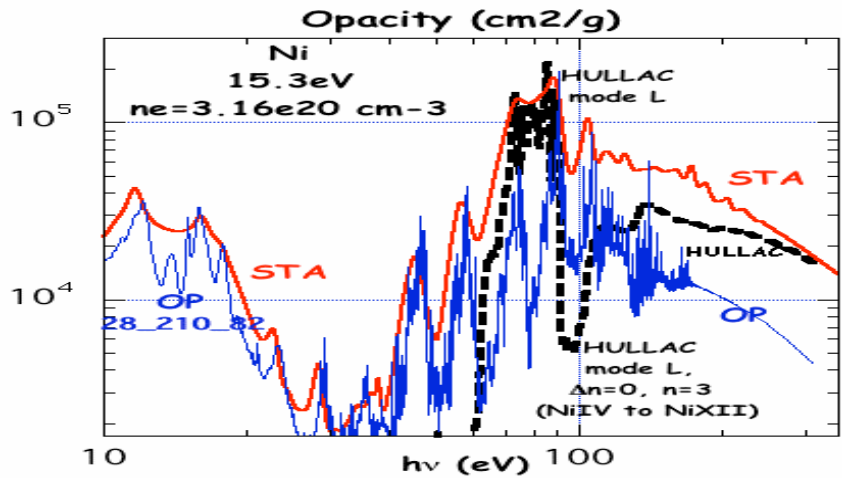

**Fig. 3.** HULLAC, OP and STA monochromatic opacity results for nickel versus energy at 15.3 eV.